\documentclass[aps,prl,reprint,a4paper,showpacs,superscriptaddress]{revtex4-1}

\usepackage{graphicx}
\usepackage{color}
\usepackage[intlimits]{amsmath}
\usepackage{amssymb}
\usepackage{amsthm}

\newcommand{\ud}{\,\mathrm{d}}
\DeclareMathAlphabet{\mathcal}{OMS}{cmsy}{m}{n}

\begin{document}

\author{G\"ulis Zengin}
\thanks{These authors contributed equally to this work.}
\affiliation{Department of Applied Physics, Chalmers University of Technology, 412 96 G\"oteborg, Sweden}
\author{Martin Wers\"all}
\thanks{These authors contributed equally to this work.}
\affiliation{Department of Applied Physics, Chalmers University of Technology, 412 96 G\"oteborg, Sweden}
\author{Sara Nilsson}
\affiliation{Department of Applied Physics, Chalmers University of Technology, 412 96 G\"oteborg, Sweden}
\author{Tomasz J. Antosiewicz}
\affiliation{Department of Applied Physics, Chalmers University of Technology, 412 96 G\"oteborg, Sweden}
\affiliation{Centre of New Technologies, University of Warsaw, Banacha 2c, 02-097 Warszawa, Poland}
\author{Mikael K\"all}
\affiliation{Department of Applied Physics, Chalmers University of Technology, 412 96 G\"oteborg, Sweden}
\author{Timur Shegai}
\email{timurs@chalmers.se}
\affiliation{Department of Applied Physics, Chalmers University of Technology, 412 96 G\"oteborg, Sweden}

\title{Realizing strong light-matter interactions between single nanoparticle 
plasmons and molecular excitons at ambient conditions}

\date{\today}

\begin{abstract}
Realizing strong light-matter interactions between individual 2-level systems and resonating cavities in atomic and solid state systems opens up possibilities to study optical nonlinearities on a single photon level, which can be useful for future quantum information processing networks. However, these efforts have been hampered by unfavorable experimental conditions, such as cryogenic temperatures and ultrahigh vacuum, required to study such systems and phenomena. Although several attempts to realize strong light-matter interactions at room-temperature using plasmon resonances have been made, successful realizations on the single nanoparticle level are still lacking. Here, we demonstrate strong coupling between plasmons confined within a single silver nanoprism and excitons in molecular J-aggregates at ambient conditions. Our findings show that deep subwavelength mode volumes, $V$, together with quality factors, $Q$, that are reasonably high for plasmonic nanostructures result in a strong coupling 
figure-of-merit --  $Q/\sqrt{V}$ as high as $\sim6\times10^{3}$~$\mu$m$^{-3/2}$, a value comparable to state-of-art photonic crystal and microring resonator cavities. This suggests that plasmonic nanocavities, and specifically silver nanoprisms, can be used for room-temperature quantum optics.
\end{abstract}


\maketitle

Strong light-matter interactions are not only interesting from a fundamental quantum optics point of view, e.g. for studying entanglement and decoherence, but also because of their relevance for high-end emerging applications such as quantum cryptography~\cite{Sci_283_2050_lo}, quantum networks~\cite{Nature_508_241_tiecke}, single atom lasers~\cite{Nature_425_268_kimble}, ultrafast single photon switches~\cite{NatPhoton_6_605_volz} and quantum information processing~\cite{PRL_83_4204_small, NatPhys_2_81_khitrova, NatPhoton_8_685_lukin}. These phenomena rely on a quantum emitter strongly interacting with a resonant cavity, which leads to cavity and emitter mode hybridization and vacuum Rabi splitting~\cite{Nature_445_896_hennessy}. In the time domain, these strong light-matter interactions manifest themselves as a coherent exchange of energy between the cavity and the emitter occurring on timescales faster than both cavity and emitter dissipative dynamics -- a situation that is dramatically different from 
irreversible spontaneous emission. Traditionally this kind of quantum optical phenomena have been studied in atomic~\cite{PRL_68_1132_kimble, Nature_443_671_aoki} and solid state systems~\cite{Nature_432_200_yoshie, Nature_446_871_zeilinger, NatPhys_4_859_faraon}, which are associated with considerable experimental challenges such as ultrahigh vacuum, cryogenic temperatures, and fabrication issues.

A possible solution to these challenges could be to use noble metal nanoparticles instead of photonic crystal and microring resonator cavities \cite{OE_18_23633_wu, ACSNano_4_6369_savasta, NL_11_2318_nordlander, PRL_110_153605_agio, ACSPhoton_1_454_tja}. This is because metal nanostructures can trap electromagnetic fields on subwavelength scales as so-called surface plasmon excitations. These plasmonic nanocavities possess a number of desirable properties, such as room temperature operation, deep subwavelength mode volumes and nanoscale dimensions that have been shown to lead to many remarkable phenomena including single-molecule Raman spectroscopy \cite{PRL_78_1667_feld, Sci_275_1102_nie, PRL_83_4357_xu}, tip-enhanced imaging \cite{Nature_498_82_zhang}, ultracompact nanolasers \cite{Nature_460_1110_noginov} and enhanced-fluorescence \cite{NatPhoton_3_654_kinkhabwala} to name a few. Despite these progresses, there has been considerably less success 
in achieving and demonstrating light-matter interactions in the strong coupling regime in these structures.

Most experiments aiming at realization of strong coupling utilize electronic excitations in a special kind of dye molecule aggregates, so-called J-aggregates, motivated by their exceptionally high oscillator strength and narrow resonances even at room temperature \cite{AngChem_50_3376_wurthner}. Recent studies along these lines include a variety of macroscopic or ensemble type systems such as propagating surface plasmons in thin metal films, low-Q Fabry-P\'erot resonators and various nanoparticle arrays or assemblies coupled to a large number of excitons \cite{PRL_93_036404_bellessa, PRB_71_035424_ebbesen, NatPhot_7_128_lienau, PRL_106_196405_ebbesen, OL_38_4498_balci}, as well as single nanoparticle measurements \cite{JPCC_111_1549_uwada, NatMeth_4_1015_liu, NL_10_77_apell, NL_13_3281_nordlander, SciRep_3_3074_gulis}. However, to date there remain ambiguities in interpretation of the plasmon-exciton interactions in these structures \cite{ACSPhoton_1_454_tja}, which from the quantum optics perspective 
translates into uncertainty in the number of excitons involved in the coupling process as well as into a question of whether plasmonic nanocavities are at all capable of realizing strong light-matter interactions at its fundamental limit. These questions are crucial for potential quantum optics applications, since those require involvement of only a single exciton \cite{Nature_445_896_hennessy, Nature_446_871_zeilinger, NatPhys_4_859_faraon}.

Here we realized light-matter interaction in the strong coupling regime between plasmons confined within single isolated silver nanoprisms and molecular excitons in J-aggregates at ambient conditions (Fig.~1). Our observations were facilitated by the weakly radiating nature of the silver nanoprisms and their small mode volumes ($V=1-7\times10^{4}$~nm$^3$). We found that the plasmon-exciton systems in our study exhibit $Q$-factors up to $\sim$20, splitting-to-damping ratios ($2g/\gamma_{pl}$) as high as $\sim$1.5 and vacuum Rabi splitting up to $\sim$280~meV involving $N\sim70$-85 excitons in the mode volume. Furthermore, our morphological and spectral measurements and analysis of a wide range of samples reveal complex spectral features as a result of diversity of silver nanoprisms and inhomogeneous distribution of J-aggregates around the nanostructures, including almost 100\% transparency dips and distinct Fano shaped scattering spectra.

\begin{figure}
\centering
\includegraphics[width=8.5cm]{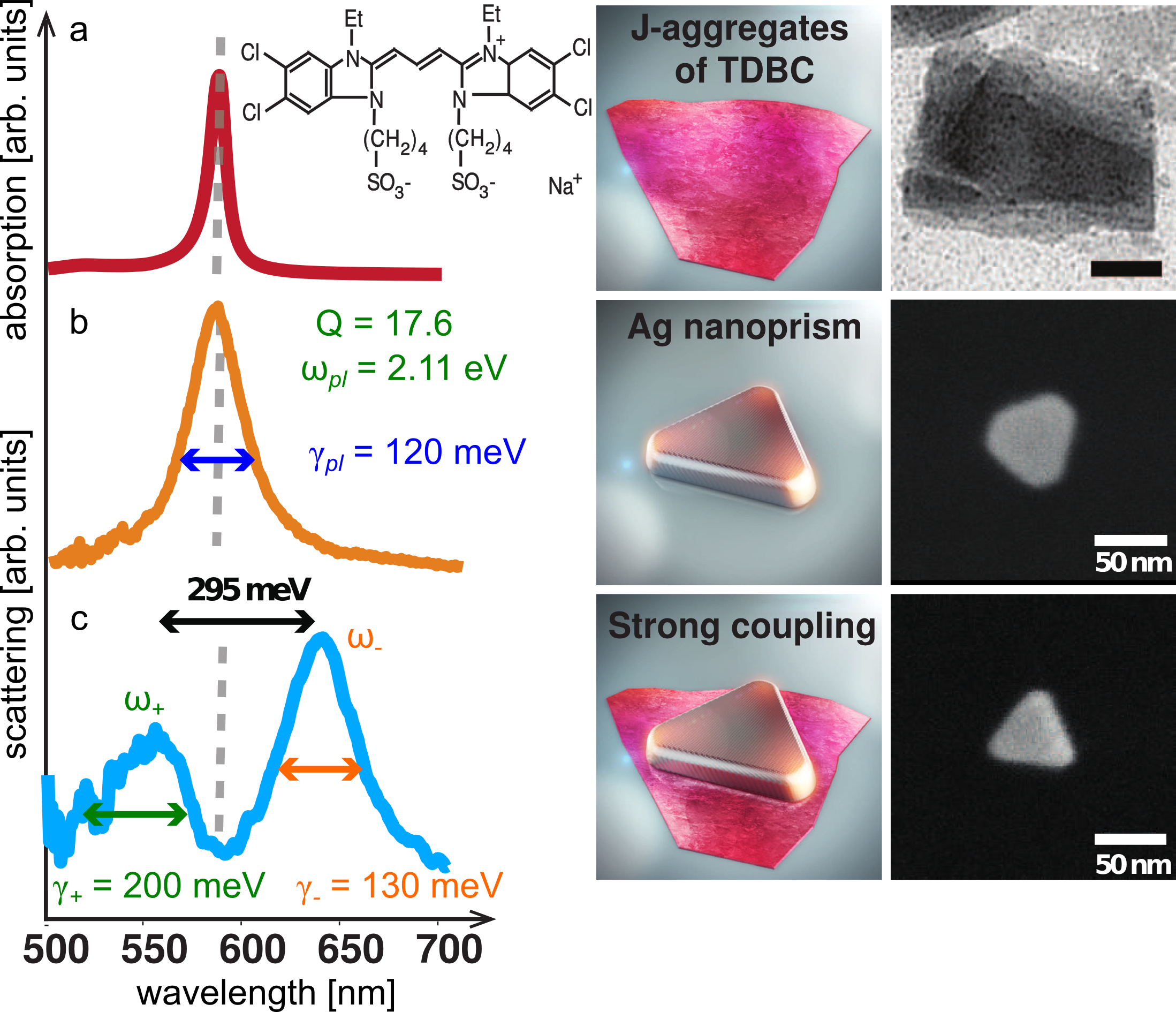}
\caption{(color online) Molecular excitons, nanoparticle plasmons and coupling between them monitored by absorption/scattering spectroscopy and electron microscopy. (a) Left: Absorption spectrum of TDBC J-aggregates in water and chemical structure of TDBC monomer. Middle: Schematic view of a J-aggregate sheet. Right: Cryo-TEM image of several overlapped J-aggregate sheets. (b) Left: Scattering spectrum of a bare nanoprism with a quality factor of 17.6. Middle: Schematic view of a silver nanoprism. Right: SEM image of the prism. (c) Left: Scattering spectrum of single nanoprism strongly coupled to J-aggregates, resulting in a pronounced scattering dip. The hybrid plasmon-exciton branches, $\omega_{+}$ and $\omega_{-}$, are split by 295~meV, while their full widths at half maxima are 200~meV and 130~meV respectively. Middle: Schematic view of the coupled system. Right: SEM image of the corresponding prism. Scale bars are 50~nm.
}
\label{fig::1}
\end{figure}

To achieve strong coupling between a cavity and an emitter, it is essential to combine a high-$Q$/low-$V$  cavity with a high oscillator strength of the emitter. For this reason, we specifically chose silver nanoprisms and J-aggregates, since the former possess low damping due to small geometrical volumes and high crystalline quality, while the transition dipole moment $\mu_{J}$ of the latter is large due to a single electron excitation being delocalized over many adjacent molecules -- all contributing to the oscillator strength \cite{AngChem_50_3376_wurthner}. The specific molecule we use is a cyanine dye called TDBC that forms J-aggregates in water solution with a single exciton delocalized over $\sim$15 molecules at room temperature \cite{JCP_102_20_koos}. Figure~1a shows the extinction spectrum of J-aggregates exhibiting a narrow peak at 588 nm ($\omega_0=2.11$~eV). When a silver nanoprism supporting a plasmon resonance matching the J-band (Fig.~1b) is placed in close 
proximity to the molecules, a coupled hybrid system exhibiting very significant mode splitting into upper ($\omega_{+}$) and lower ($\omega_{-}$) hybrid plasmon-exciton branches is formed, signaling the realization of a strong coupling scenario (Fig.~1c). A possible dye-nanoparticle arrangement that could lead to such scattering spectrum is schematically shown in Fig.~1 -- as supported by scanning and transmission electron (SEM and cryo-TEM) and atomic force (AFM) microscopies showing that J-aggregates form thin $\sim$3~nm planar sheets and silver nanoprisms reside above or below them (Fig.~S1).

Figure~1 shows that silver nanoprisms can indeed strongly interact with J-aggregates. To reveal the factors affecting the coupling process, we measured and analyzed scattering spectra from $\sim$20 isolated plasmon-exciton systems, structurally characterized them and performed electrodynamics finite-difference time domain (FDTD) calculations. Individual silver nanoprisms' spectra were measured using hyperspectral imaging -- an approach that allows for parallel sampling of many isolated particles \cite{JPCC_113_16839_vanDuyne, NL_11_1826_kall} (Fig. S3). Control experiments ensuring that the scattering dips indeed originate from plasmon-exciton coupling were performed: monitoring spectral evolution as a function of J-aggregate photodecomposition (Fig.~3), fluorescence of hybrid systems (Fig.~S8) and bare silver nanoparticle scattering measurements (Fig.~S4, S9). 

To determine whether the plasmon-exciton system is strongly coupled, we describe our scattering spectra in terms of the classical coupled harmonic oscillator model, which predicts upper and lower plasmon-exciton branches in agreement with the quantum mechanical Jaynes-Cummings picture \cite{PRL_93_036404_bellessa, PRB_59_10227_rudin, PRB_67_085311_agranovich}:
\begin{equation}
\omega_{\pm}=\frac{1}{2}\left(\omega_{pl}+\omega_0\right)\pm\sqrt{g^2 + \frac{\delta^2}{4}},
\end{equation}
Here $g$ is the coupling rate, $\omega_{pl}$ and $\omega_0$  are plasmon and exciton resonance energies and $\delta=\omega_{pl}-\omega_0$  is the detuning. The plasmon-exciton branches can be directly accessed from the scattering data (Fig.~1c). Assuming the exciton resonance and width are homogeneous over the whole set of experiments ($\omega_0\approx2.11$~eV and $\gamma_0\approx100$~meV), we obtain the vacuum Rabi splitting  $\Omega_{R}=2g$, plasmon resonance and plasmon linewidth from Eq.~(1) as  $\Omega_R=2\sqrt{(\omega_{+}-\omega_0)(\omega_{0}-\omega_{-})}$, $\omega_{pl}=\omega_{+}+\omega_{-}-\omega_0$ and $\gamma_{pl}=\gamma_{+}+\gamma_{-}-\gamma_0$ , where $\gamma_{\pm}$  is the full width at half maximum of the corresponding plasmon-exciton branch. The quality factor is calculated as  $Q=\frac{\omega_{pl}}{\gamma_{pl}}$. The coupled oscillator model is an alternative to direct determination of $\Omega_{R}$ via anti-crossing at zero detuning, which is typically used in case of e.g. tunable photonic 
crystal cavities \cite{NatPhys_2_81_khitrova}. Here the nanocavity resonances are not tunable, however, by looking at the upper and lower plasmon-exciton branches for all measured nanoprisms (Fig.~S5), we observe that the resulting curves resemble an anti-crossing behavior with $\Omega_{R}$ of $\sim$170 meV, in good agreement with Fig.~2b. The analysis reveals that $\omega_{pl}$, $\gamma_{pl}$, $Q$ and $\Omega_{R}$ of the coupled plasmon-exciton systems are distributed across a broad range of values (Fig.~2b-e), reflecting variations in particle morphology (Fig.~2f) as well as in the number of excitons contributing to the coupling process, $N$. Since $g=\sqrt{N}\mu_J|E_{vac}|$, where $|E_{vac}|=\sqrt{\frac{\hbar\omega}{2\epsilon\epsilon_{0}V}}$ is the vacuum field and $V$ is the mode volume \cite{Nature_432_200_yoshie}, we should comment on the physical meaning of $V$ for plasmonic cavities. Following \cite{OL_35_4208_koenderink, ACSPhoton_1_2_kristensen}, we note that in case of small plasmonic 
nanoparticlesis the mode volume is well approximated by the geometrical volume, becasue the energy density -- $\epsilon|E|^2$ is concentrated mostly in the metal (Fig.~2a inset). This implies that $V$ is given by the nanoprism side lengh, $L$, as $V\sim L^2$ since the nanoprism's height is constant 10~nm. Thus, the coupling process can be seen as given by only two parameters -- $N$ and $L$. 

\begin{figure}
\centering
\includegraphics[width=8.5cm]{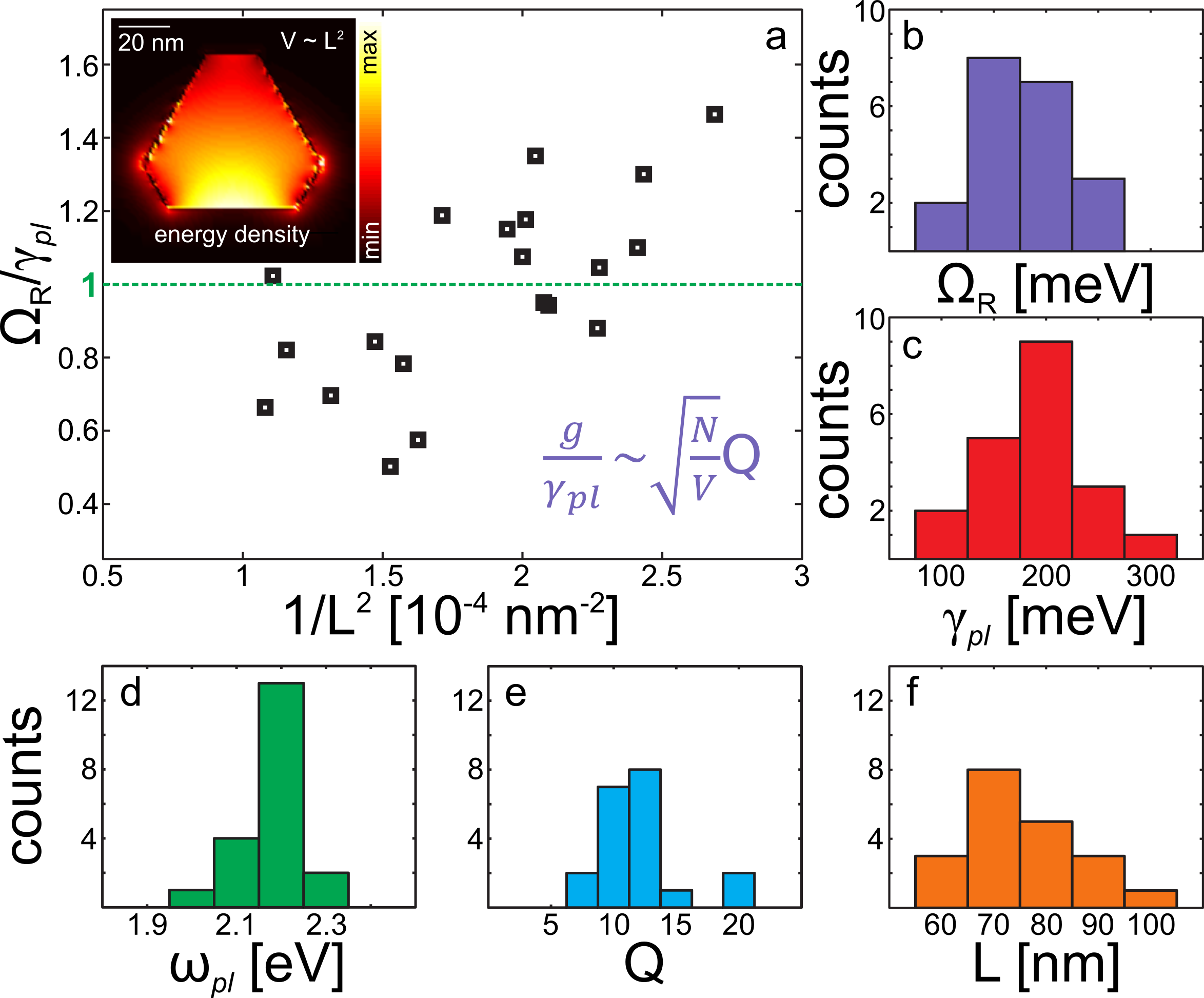}
\caption{(color online) Distribution of single nanoparticle parameters extracted using the coupled oscillator model. (a) Ratio between Rabi splitting and plasmon damping ($\frac{2g}{\gamma_{pl}}$) as a function of inverted geometrical volume ($L^{-2}$). Strong coupling is reached when $2g>\gamma_{pl}$  is fulfilled. Inset: energy density $\epsilon|E|^2$ distribution around a silver nanoprism, showing that $V$ is well approximated by geometrical volume, i.e. $V\sim L^2$. (b) Distributions of Rabi splitting $\Omega_R$, (c) plasmon damping  $\gamma_{pl}$, (d) plasmon resonance frequencies  $\omega_{pl}$, (e) quality factors, and (f) nanoprisms' side lengths. }
\label{fig::2}
\end{figure}

For further analysis we should first confine ourselves to a specific criterion of strong coupling taken as $2g>\gamma_{pl}$  in this study. This is a rather strict criterion in comparison to $2g>(\gamma_{pl}-\gamma_0)/2$  or $2g>(\gamma_{pl}+\gamma_0)/2$  used in other works \cite{NatPhys_2_81_khitrova, NL_13_3281_nordlander} and therefore satisfaction of this strict condition automatically implies satisfaction of all the weaker ones. The parameters extracted from the coupled oscillator model -- $2g$, $\omega_{pl}$ and $\gamma_{pl}$ -- on one hand allow to infer whether the system satisfies the strong coupling condition and on the other to estimate $N$ and $V$ through the standard relation for the coupling rate $g/\gamma_{pl}\sim Q\sqrt{N/V}$  \cite{Nature_432_200_yoshie}. That said, in Fig.~2a we plot $2g/\gamma_{pl}$  as a function of $L^{-2}$ (mimicking dependence on $V$) and observe that several plasmon-exciton systems exhibit strong 
coupling, i.e. $2g/\gamma_{pl}>1$. We also observe that $2g/\gamma_{pl}$ increases for smaller particles, however, the correlation is not very strong, which is likely due to the diversity of the nanoprisms and inhomogeneous distribution of molecules around them. To verify whether such spread is realistic, we performed electrodynamic FDTD calculations (Fig.~S6), which show that silver nanoprisms indeed can be expected to have  $Q$ values in the range of 8-14 and the mode volume in the range of $V=1-7\times10^4$~nm$^3$ in good correlation with geometrical volume. Such mode volume dispersion agrees well with the $2g/\gamma_{pl}$  spread shown in Fig.~2a. We would like to stress here that a quantity $Q/\sqrt{V}$, which characterizes the cavity in terms of its coupling ability and therefore referred to in the literature as the figure-of-merit for strong coupling related phenomena \cite{Nature_432_200_yoshie, OpEx_13_5961_englund}, is about $\sim6.3\times10^{3}$~$\mu$m$^{-3/2}$ in case of silver nanoprisms studied 
here 
($Q=20$ and  $V=10000$~nm$^{3}=10^{-5}$~$\mu$m$^3$), which is only 5 times smaller than state-of-art photonic crystal cavities $\sim3.2\times10^4$~$\mu$m$^{-3/2}$ ($Q=10000$ and $V=0.1$~$\mu$m$^3$ -- parameters taken from Ref. \cite{Nature_432_200_yoshie}). This implies that plasmonic nanoparticles, especially in the form of weakly radiating single crystalline nanoprisms studied here, are indeed very prominent alternatives to photonic crystal and microring resonator cavities.

To elucidate how many excitons contribute to the observed interactions, we used the experimentally obtained values for $g$ together with numerically evaluated $V$, and compared them using the standard relations for the coupling rate and the vacuum field. For the two limiting cases representing the data ($g=140$~meV, $V=10000$~nm$^3$ and $g=50$~meV, $V=70000$~nm$^3$), we estimate $\sqrt{N}\mu_{J}=\frac{g}{|E_{vac}|}\approx170-190$~Debye. $\mu_J$ is independently evaluated from the extinction measurement in water solution of J-aggregates $\approx20.5$~Debye (Fig.~S2) and thus the overall number of excitons contributing to $g$ is $\approx70-85$, while the number of TDBC monomers is $\sim1000$. Although, the current realization is rather far from the quantum optics limit, by extrapolating the coupling down to a single exciton we obtain $g=15$~meV, which can be realistically increased by further compressing the mode volume and increasing $\mu_J$.

The plasmon-exciton system used in our study is amenable for realizing a wide range of light-matter interactions (Fig.~2a). To illustrate this diversity we have chosen and characterized five of the nanoprisms in greater detail. The realized coupling regimes range from very strong to weak. Figure~3a shows a strongly coupled plasmon-exciton system with a Rabi splitting of $162$~meV, while $\gamma_{pl}=109$~meV and $Q\approx20.2$, resulting in a value of $\frac{2g}{\gamma_{pl}}$  of about 1.5 -- a very high number for both plasmonic and photonic systems \cite{NatPhys_2_81_khitrova, NL_13_3281_nordlander, SciRep_3_3074_gulis}. Importantly, Fig.~3a shows a system with relatively modest Rabi splitting of $\sim162$~meV, while the maximum observed in this study is $\sim280$~meV, yet it is deep in the strong coupling regime -- achieved due to relatively high $Q$ in this case. In Fig.~3b we show the scattering spectrum for a nanoprism resembling a case of intermediate coupling ($\gamma_{pl}>2g>\gamma_0$). Here, 
$2g=151$~meV, while $\gamma_{pl}=184$~meV and  $Q\approx11.6$. The characteristic of this regime is that the dip is not as pronounced as in the strong coupling case -- a result of increased $\gamma_{pl}$ and an average $Q$. However, coupling in the intermediate regime is still more pronounced than in the weak coupling case shown in Fig.~3c, where the scattering spectrum is only slightly suppressed at the position of the J-aggregate line -- a situation realized when splitting is low, i.e. $\gamma_{pl}\gg g$  ($\gamma_{pl}\approx191$~meV, $\Omega_{R}\approx96$~meV). Moreover, it has been shown previously that suppression of scattering in case of weak coupling occurs mostly due to enhanced absorption in the dye layer \cite{ACSPhoton_1_454_tja}. A situation observed in Fig.~3c is similar to several previously published single-particle data \cite{JPCC_111_1549_uwada, NatMeth_4_1015_liu, NL_10_77_apell}, suggesting realization of an enhanced-absorption scenario in these works.

\begin{figure}
\centering
\includegraphics[width=8.5cm]{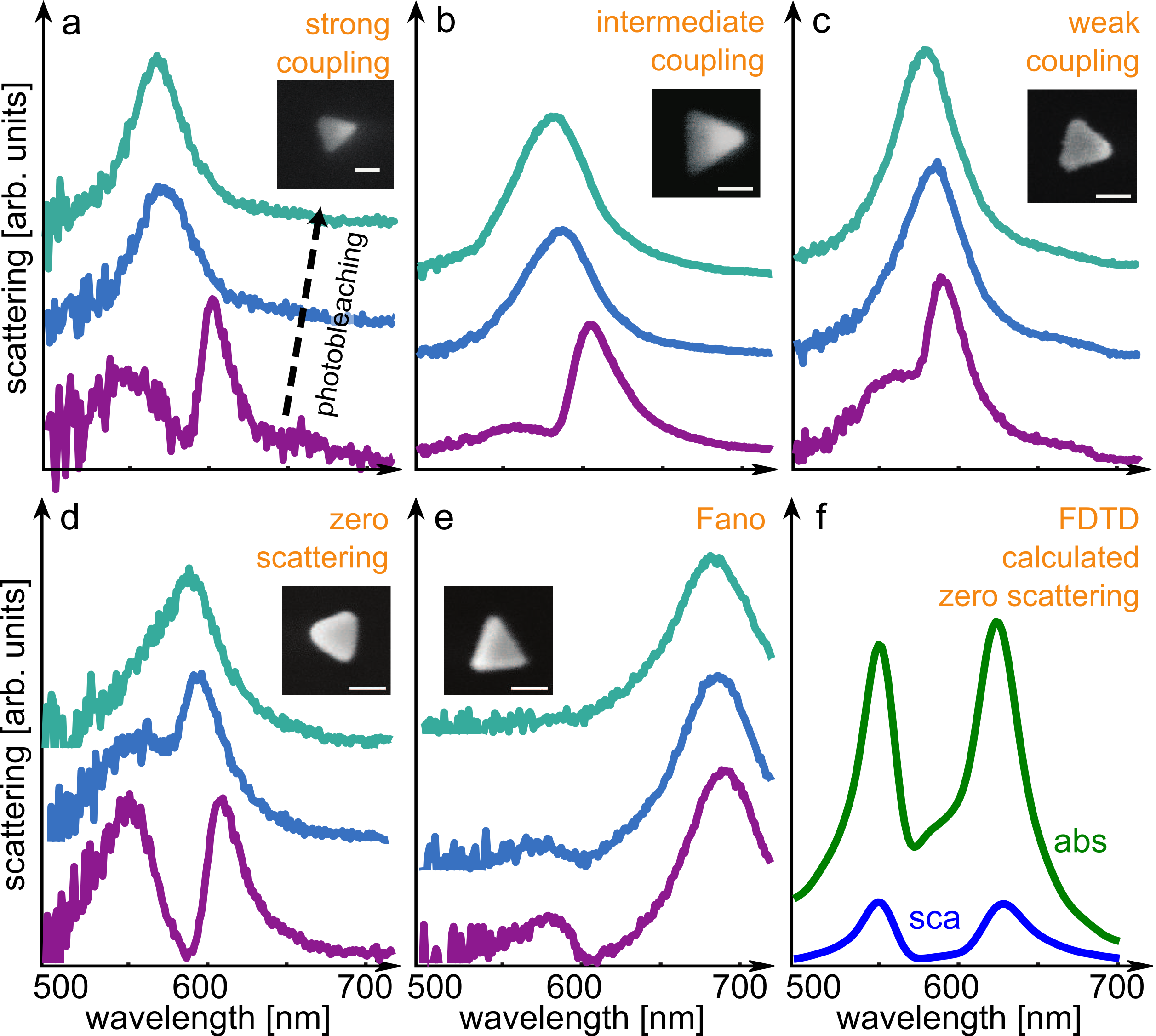}
\caption{(color online) Diversity of plasmon-exciton interactions. (a) A single nanoparticle--J-aggregate system shown together with laser-induced J-aggregate photodegradation 2 and 10 minutes after continuous laser illumination for strongly coupled $2g>(\gamma_{pl},\gamma_{0})$, (b) intermediate coupling -- $\gamma_{pl}>2g>\gamma_0$ and (c) weak coupling regimes  $2g\ll\gamma_{pl}$. (d) Near-complete suppression of scattering (zero-scattering) at around the J-band (588 nm) in the strong coupling regime. (e) Fano-shaped scattering spectrum. SEM images of the nanoparticles are shown in insets. The scale bars are 50 nm. (f) FDTD calculated absorption and scattering, note the splitting of both cross sections.}
\label{fig::3}
\end{figure}

Interestingly, the spectrum in Fig.~3d exhibits a nearly complete suppression of scattering in the strong coupling regime (i.e. zero-scattering), with $2g\approx200$~meV dominating over $\gamma_{pl}\approx174$~meV. From Ref. \cite{OE_18_23633_wu, SciRep_3_3074_gulis}, it can be shown that very pronounced dips in scattering response can be achieved when $\frac{2g}{\sqrt{\gamma_0\gamma_{pl}}}\gg1$, which is true for both Fig.~1c and Figs.~3a,d, suggesting that strong coupling is accompanied not only by large splitting but also by nearly complete suppression of scattering. The phenomenon of zero scattering requires that the absorption cross-section of these nanostructures should be also strongly suppressed, in accordance with the optical theorem. 
To illustrate this, we perform FDTD caclulations using realistic nanoprism-J-aggregate parameters (Fig.~3f), which indicate that spectral dips are present not only in scattering but also in absorption -- as anticipated in case of strong coupling \cite{ACSPhoton_1_454_tja}.

Finally, in Fig.~3e, we observe an asymmetric Fano lineshape in the plasmon-exciton system, arising due to interaction between narrow J-aggregate resonance and a broad detuned plasmon resonance ($\omega_{pl}\approx1.83\text{ eV}\ll\omega_0$). This example aims at illustrating similarities between various aspects of plasmon related phenomena often referred as Fano resonances, strong coupling or electromagnetically induced transparency.

In all five cases presented in Fig.~3, independently of the realized scenario, the scattering dips gradually disappeared upon laser-induced J-aggregate degradation, implying that the diversity arises directly from plasmon-exciton interactions. Indeed, SEM images confirm that in each case a spectrum originated from an isolated nanoprism. Fluorescence spectra of these plasmon-exciton systems (Fig.~S8) are much broader than fluorescence of the free J-aggregates in solution, indicating strong interaction between J-aggregates and surface plasmons. Note that control experiments on bare silver nanoprisms showed that the laser illumination does not affect the nanoparticles' shape and spectra (Fig.~S9).

In conclusion, strong light-matter interactions realized in the single crystal nanoprisms studied here outperform all previously reported realizations, including Ag nanorods \cite{SciRep_3_3074_gulis} and Au dimers \cite{NL_13_3281_nordlander} in terms of degree of coupling -- $2g/\gamma_{pl}$ --  and strong coupling figure-of-merit $Q/\sqrt{V}$. We note that $2g/\gamma_{pl}$  and $Q/\sqrt{V}$  are more relevant for the description of strong coupling than the vacuum Rabi splitting alone, no matter how high the latter is. Indeed,  ensemble measurements reported Rabi splitting approaching $\sim1$~eV \cite{CPC_14_125_ebbesen} -- more than three times greater than in the current study, however, this was achieved due to tremendous amount of molecules adsorbed within large mode volumes of macroscopic samples. Vacuum Rabi splitting in the range 200-400~meV was also claimed for the case of Au dimers produced by electron beam lithography \cite{NL_13_3281_nordlander}, however, these 
polycrystalline dimer structures exhibit significant radiative losses resulting in $\gamma_{pl}$ as large as $\sim370$~meV, thereby reducing  $2g/\gamma_{pl}$ to values of about $\sim1.08$ even for the only nanostructure displaying $\sim400$~meV Rabi splitting. In the current study, $\gamma_{pl}$  is strongly suppressed due to reduced radiative damping and high crystalline quality of silver nanoprisms resulting in $2g/\gamma_{pl}\approx1.5$  and $Q/\sqrt{V}\approx6\times10^{3}$~$\mu$m$^{-3/2}$ as demonstrated by both experiments and calculations. These results allow us to estimate the number of excitons coherently contributing to the coupling process, $N\sim70-85$, implying that single exciton strong coupling might be within reach provided further compression of the mode volume. Furthermore, photobleaching and structural characterization experiments unambiguously prove that coupling arises due to plasmon-exciton interactions. Importantly, our results show that plasmon-exciton systems are viable alternatives 
to photonic cavities, thereby opening exciting opportunities for room temperature quantum optics.

We would like to thank Andr\'e Dankert for helping with AFM measurements, Dr. Annika Altsk\"ar for helping with cryo-TEM measurements and Dr. Stefan Gustafsson for helping with TEM measurements. The authors acknowledge financial support from Swedish Research Council (VR), Knut and Alice Wallenberg Foundation (KAW), Swedish Foundation for Strategic Research (SSF), and the Foundation for Polish Science via the project HOMING PLUS/2013-7/1.


%

\newpage

\renewcommand{\theequation}{S\arabic{equation}}
\renewcommand{\thefigure}{S\arabic{figure}}
\setcounter{figure}{0}
\setcounter{equation}{0}

\begin{widetext}

{\centering \large{\textbf{Supplemental Material for \\``Realizing strong light-matter interactions between single nanoparticle plasmons and molecular excitons at ambient conditions"}}

\normalsize{G\"ulis Zengin, Martin Wers\"all, Sara Nilsson, Tomasz J. Antosiewicz, Mikael K\"all, and Timur Shegai

\emph{Department of Applied Physics, Chalmers University of Technology, 412 06 G\"oteborg, Sweden}

\emph{Centre of New Technologies, University of Warsaw, Banacha 2c, 02-097 Warszawa, Poland}}

}

\end{widetext}

\tableofcontents

\section{Methods}

\subparagraph{Sample preparation}

The Ag colloidal particles were synthesized by the light-induced ripening process \cite{Sci_294_1901_jin}. TDBC dye (5,6-Dichloro-2- [[5,6-dichloro-1-ethyl-3-(4-sulfobutyl)-benzimidazol-2-ylidene]-propenyl]-1-ethyl-3-(4-sulfobutyl)-benzimidazolium hydroxide, inner salt, sodium salt) was purchased from FEW Chemicals. The formation of J-aggregates from monomers depends on a number of factors including dye concentration, ionic environment and pH. By optimizing these parameters, the aqueous solution of TDBC dye was prepared at concentration of $\sim10$~$\mu$M in 5~mM NaOH. 50~$\mu$l of Ag solution was mixed with 50~$\mu$l of J-aggregate solution. The mixture was let for incubation overnight. The J-aggregate/Ag nanoparticles mixture was then applied to a TEM grid precoated with polylysine.

\begin{figure*}
\centering
\includegraphics[width=15cm]{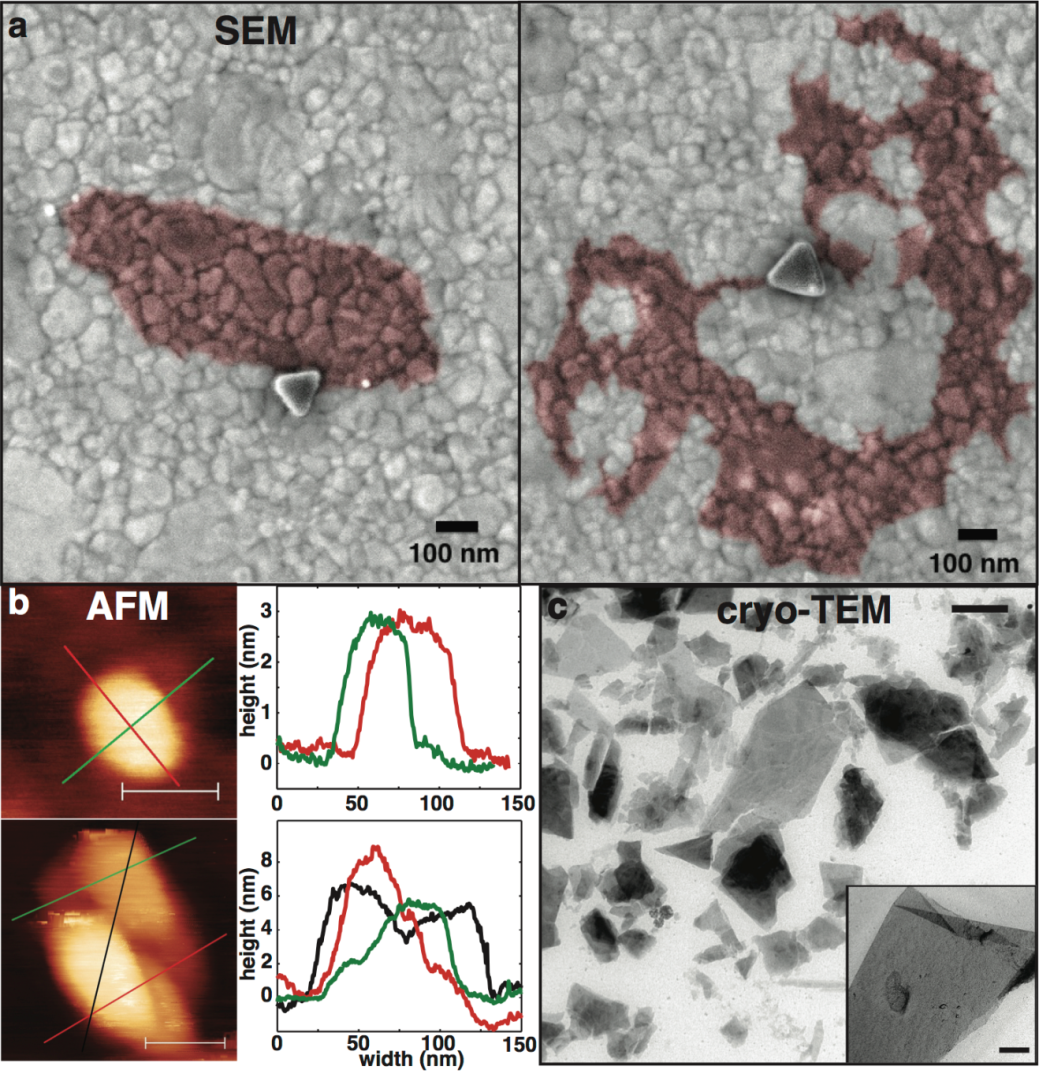}
\caption{Structural characterization of free and coupled J-aggregates. (a) SEM of Ag nanoprism -- J-aggregate (pink) composites on top of gold film (grainy background). (b) Topography of J-aggregate sheets on top of silicon substrate as recorded by AFM. Top: a single layer. Bottom: bi- and tri- layers of J-aggregates. (c) J-aggregate sheets characterized by cryo-TEM. Inset: a single folded flake.}
\label{fig::S1}
\end{figure*}

\subparagraph{Optical measurements}

Single particle dark-field scattering and fluorescence measurements were performed in an inverted microscope (Nikon TE-2000E) equipped with variable numerical aperture oil immersion objective (100X, NA = 0.5-1.3, Nikon). A tunable liquid crystal filter (VariSpec, 400-720~nm), that is transparent for light of only certain wavelength and linear polarization was used together with iXon EM-CCD detector (Andor) for hyperspectral imaging. Nearly isotropic in-plane polarization response of the nanoprisms justifies usage of the tunable filter in a single polarization channel. The transmission window of the liquid crystal filter was 10~nm. The liquid crystal filter was set to perform wavelength steps of 1~nm between successive images and synchronized with the CCD-chip. The hyperspectral images were recorded in sequences with wavelengths ranging from 500-720~nm and 535-650~nm for the dark-field and fluorescence measurements respectively. In case of triangular nanoprisms studied here, the method has an additional 
technical advantage, that is, many of the particles turned out to be aggregates of several silver prisms, thereby effectively lowering the throughput of the measurements. Only single nanoparticle data was analyzed in this study. In dark-field measurements, an air dark-field condenser (Nikon, NA~=~0.8-0.95) was used. Tungsten halogen lamp was used to illuminate the sample. Even when driven at the maximum power, the light intensity from the lamp was still low enough to ensure no significant photodegradation of the dye samples. For fluorescence measurements, the sample was illuminated with a 532~nm laser at power of 100 W/cm$^2$ in epi-illumination mode. The colored Rayleigh scattering and fluorescence images were taken by Nikon D300s DSLR camera.

\subparagraph{Electron microscopy}

High-vacuum scanning electron microscopy (SEM) with 2~kV acceleration voltage was used for imaging the single nanoparticles. As described in the sample preparation section, the particles were applied on top of a copper-grid substrate containing several distinct hollow quadrants. Each quadrant was supported with a very thin membrane on which the particles finally landed. Furthermore, in the center of the copper-grid, an asymmetric alignment mark was present which enabled us to locate and map out the quadrant of interest. Each particle inside a given quadrant was correlated with a corresponding optical image, based on which a correlated SEM-optical map was obtained.

\section{\textbf{S1}: Structural characterization of J-aggregates/Ag nanoparticle hybrids}

The results of our morphological characterization are summarized in Figure S1. The data show that J-aggregates are organized in large planar sheets of about $\sim3$~nm thickness (see AFM scan) -- somewhat similar to 2D materials e.g. graphene. Multiple layers, folded regions and defects are sometimes seen as evidenced by different level of contrast observed in both AFM and cryo-TEM images. AFM Scans along various directions (see Fig. S1b Bottom) show that the height profile varies in steps of about 2.5-3~nm, indicating occasional formation of bi-, tri- and multi-layers. The lateral size distribution of J-aggregate flakes is rather broad and ranges from about few tens of nanometers to tens of microns (see cryo-TEM). The largest of them can be easily observed in scattering.

Silver nanoparticles were found to lie flat either above or below J-aggregate sheets (see Fig. S1a). As is seen from those examples, due to very broad size distribution of J-aggregates, some of them can in principle be much larger than nanoprisms. If such a system is observed in a scattering experiment, most of the aggregate would not couple to the plasmon mode at all, but the non-interacting part would nevertheless contribute to scattering at J-band wavelength – a situation that was never observed in experiments. We therefore conclude that the optical data presented in Figs. 1-3 probably originates from J-aggregates whose lateral dimensions are comparable, but not much larger than the size of the nanoprism. These reasonably small J-aggregates are hard to directly visualize in SEM, nonetheless the structures shown in Figure S1a provide a reasonable idea of the plasmon-molecule organization. This hypothesis is additionally supported by numerical simulations using realistic dimensions and oscillator strength 
for the J-aggregate layers (see Figs. S7-S9).

\section{\textbf{S2}: Extinction of a free J-aggregate}

\begin{figure}[t]
\centering
\includegraphics[width=7.5cm]{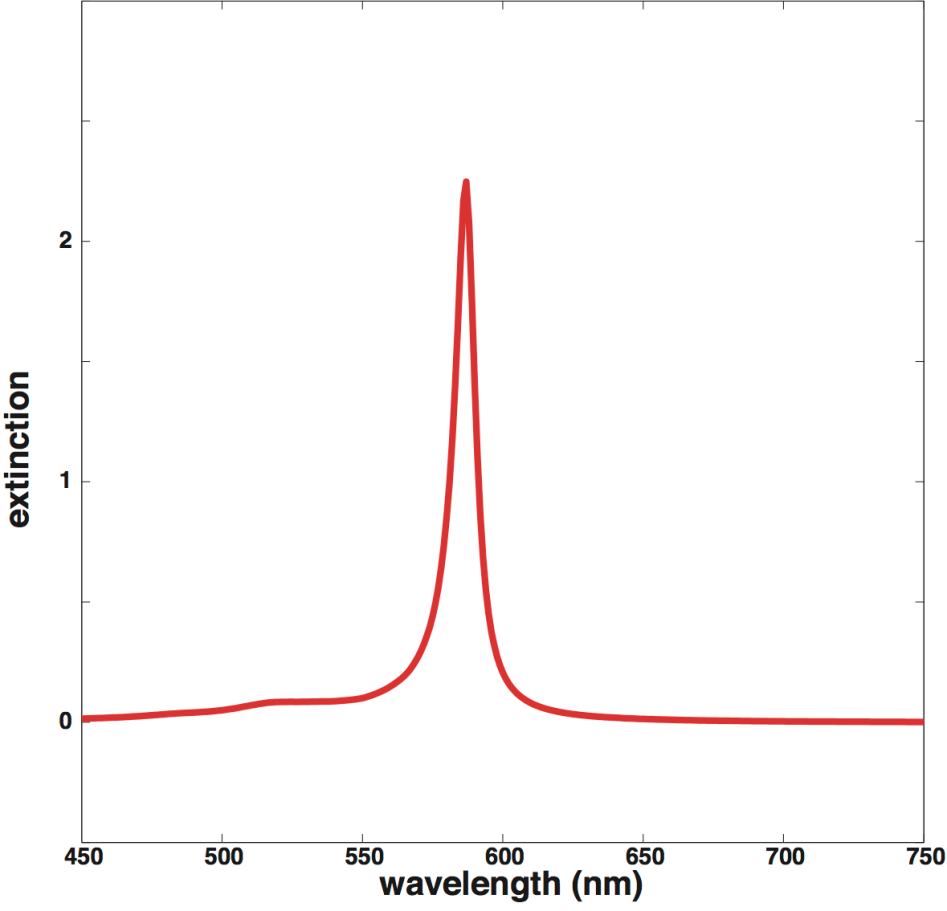}
\caption{Extinction optical density spectrum of 15~$\mu$M J-aggregate TDBC dye molecules mixed in 0.2~mM NaOH buffer solution. The measurement was conducted with the solution placed in a 0.4 cm thick quartz cuvette.}
\label{fig::S2}
\end{figure}

\subparagraph{Free dye parameters}

An extinction measurement performed on free J-aggregates in water solution can be used to evaluate   independently. In these estimations, it is important to have in mind that the exciton is strongly delocalized over many molecules forming the J-band and that each of the molecules involved contributes to the total oscillator strength of the transition. For TDBC, the literature suggests the room-temperature delocalization to occur over $N_d\approx15$ molecules, irrespectively of the actual physical size of the aggregate, which can be much larger \cite{JCP_102_20_koos}. Keeping the delocalization in mind, one can obtain the absorption cross-section of the exciton as  $\sigma_{J}\approx\frac{2.3DN_d}{n_ml}\approx2.15\times10^{-14}$~cm2, where  $D$ is the extinction optical density ($\sim2.25$ at maximum), $n_m$  is the concentration of monomers (15~$\mu$M = $9\times10^{15}$~cm$^{-3}$) and $l$  is the light path length (0.4~cm) in the extinction measurement. Transition dipole moment characterizing the delocalized 
exciton is then given by \cite{AmJPhys_50_982_hilborn}, $\mu_{J}=(\frac{3}{4\pi^2}\hbar\sigma_{J}\gamma_{0}\epsilon_{0}\lambda)^{0.5}$ , where $\gamma_{0}$  is the FWHM of the extinction spectrum (in s$^{-1}$), $\epsilon_0$  -- vacuum permittivity and $\lambda$  is the resonance wavelength. Substituting all the measured parameters, we obtain the charge displacement  $d\approx4.3$~\AA\/ and the dipole moment  $\mu_{J}=ed\approx20.5$~Debye of one exciton.

\begin{widetext}
\newpage

\section{\textbf{S3}: Hyperspectral imaging}

\begin{figure*}
\centering
\includegraphics[width=16cm]{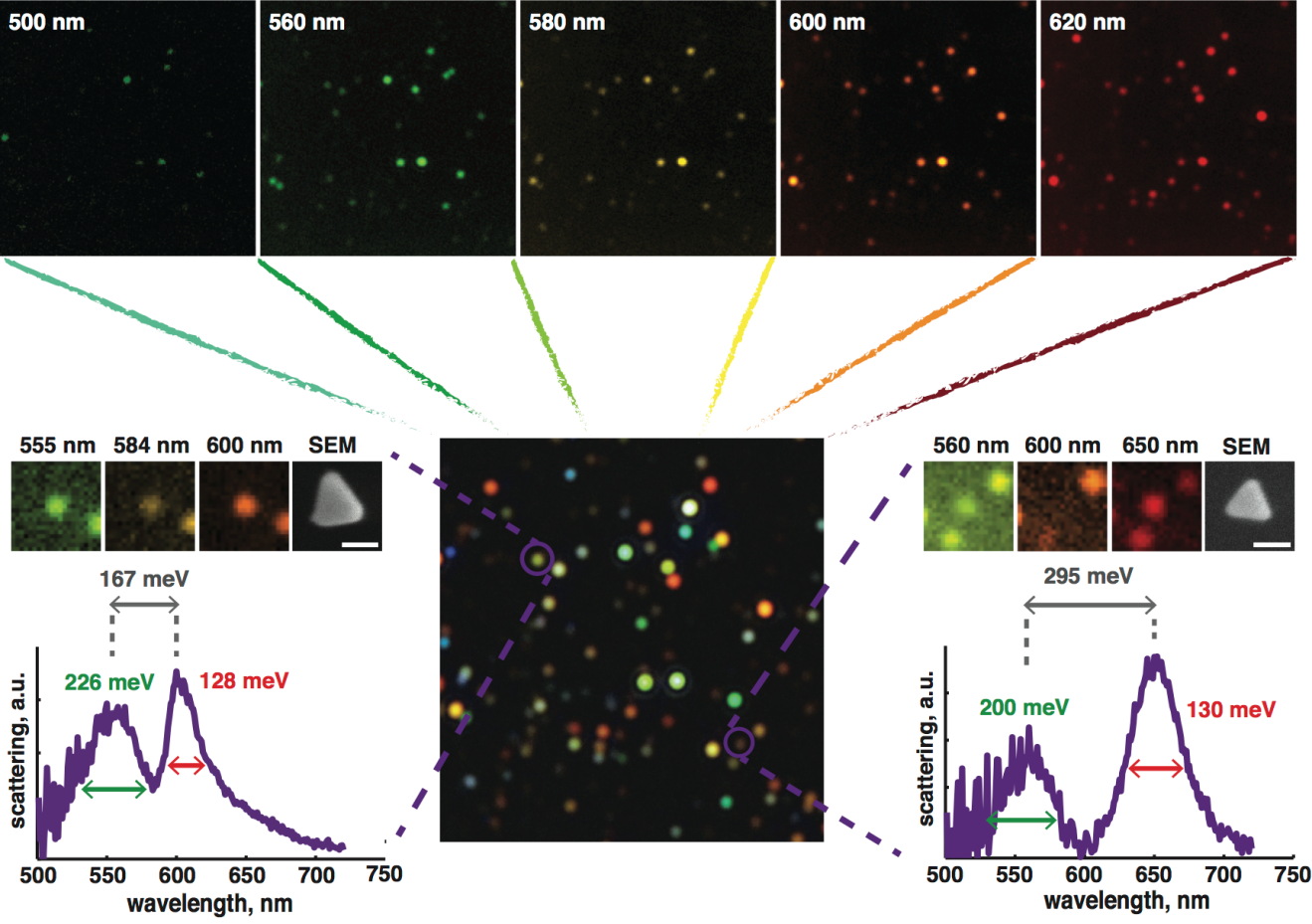}
\caption{Illustration of hyperspectral imaging for single nanoparticle measurements. The lower central image displays the actual dark-field scattering map, where each colorful spot corresponds to Rayleigh scattering emitted by a single Ag nanoparticle or aggregate. Particles (1) and (2) are marked with violet circles and their corresponding spectra are shown next to the dark-field map. SEM images and color-coded dark-field images recorded at wavelength corresponding to peaks and dips in the spectra are shown in the inset. Upper figure sequence show the change in pixel intensity for five different wavelengths using the same dark-field image with a tunable liquid crystal filter applied in front of the CCD.}
\label{fig::S3}
\end{figure*}

\end{widetext}

In this study the experiments were done by using a hyperspectral imaging technique utilizing a tunable liquid crystal filter. The concept of ``spectroscopy by imaging" is illustrated in Fig. S3. Rayleigh scattering images in dark-field configuration were recorded at a number of spectrally narrow intervals controlled by a tunable filter. The scattering spectrum was then reconstructed by monitoring the intensity of a given particle as a function of wavelength. This procedure allows for parallel sampling of many isolated particles over the entire visible range and tremendously increases the throughput of single nanoparticle spectroscopy measurements \cite{JPCC_113_16839_vanDuyne, NL_11_1826_kall}. Dark-field scattering images of silver nanoparticles/J-aggregates hybrids recorded in this way are shown in the top row of Fig. S3 (for five exemplary wavelengths 500, 560, 580, 600, and 620 nm). The images are approximately $35\times35$~$\mu$m. Two representative single nanoprism spectra together with their SEM 
images are shown. The first particle (p1) exhibits a suppressed scattering at around the J-aggregate resonance, however, the mode splitting does not overcome the plasmon resonance width. The second particle (p2), on the other hand, shows much wider splitting which clearly overwhelms both plasmon and molecular decoherence rates and therefore enters the true strong coupling regime. Vacuum Rabi splitting for the case of particle (2) reaches about 300 meV, well above the plasmon resonance width, as shown in Fig. S3. These two particles, and the brief analysis of their spectral differences, aim at illustrating the power of hyperspectral imaging for single particle spectroscopy.

\newpage

\begin{widetext}
\newpage

\section{\textbf{S4}: Bare Ag nanoprisms statistics}

\begin{figure*}[h]
\centering
\includegraphics[width=15cm]{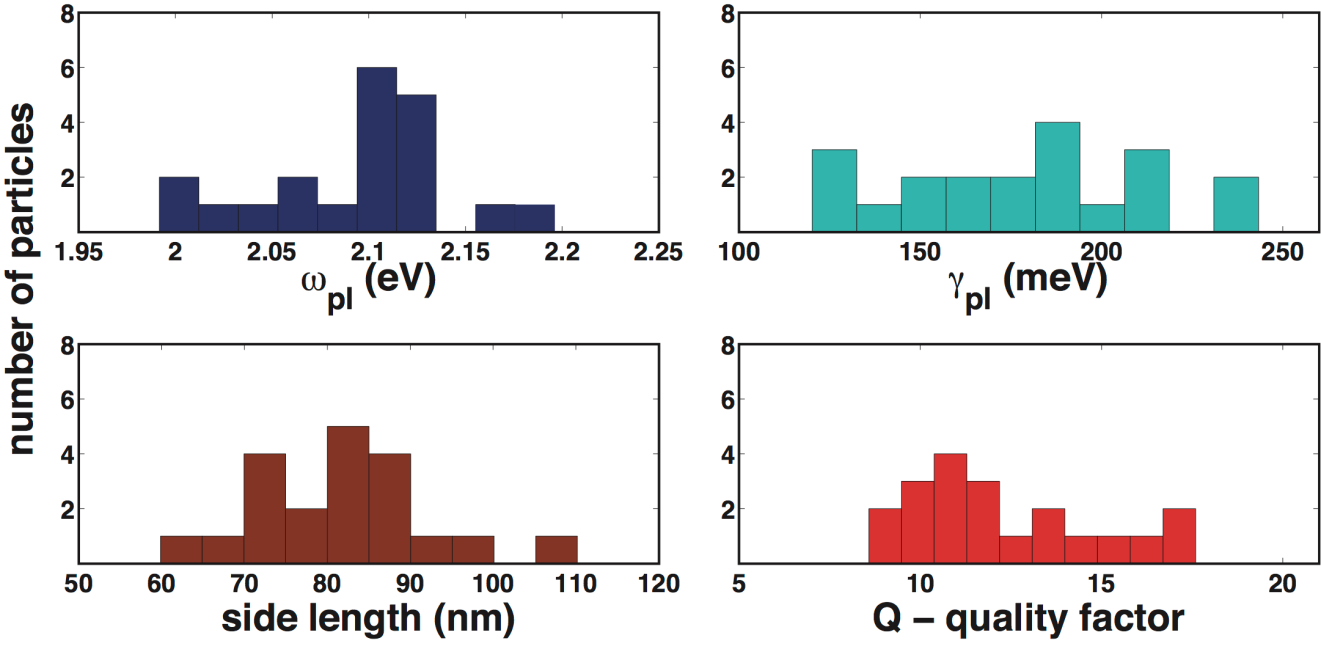}
\caption{Distribution of  $\omega_{pl}$, $\gamma_{pl}$  side lengths and quality factors for bare silver nanoprisms not interacting with J-aggregates. The values are based on optical and SEM measurements on single nanoparticle. Average values correspond to $\omega_{pl}\approx2.07$~eV, $\gamma_{pl}\approx170 $~meV, $L\approx80$~nm and $Q\approx12$, which agrees well with the average values of corresponding parameters extracted using the coupled oscillator model and shown in Fig. 2 of the main text, as well as with SEM data for coupled nanoprisms shown in Fig. 2f.}
\label{fig::S4}
\end{figure*}

\section{\textbf{S5}: Mode anti-crossing and scattering data for all nanoparticles}

\begin{figure}[h]
\centering
\includegraphics[width=15cm]{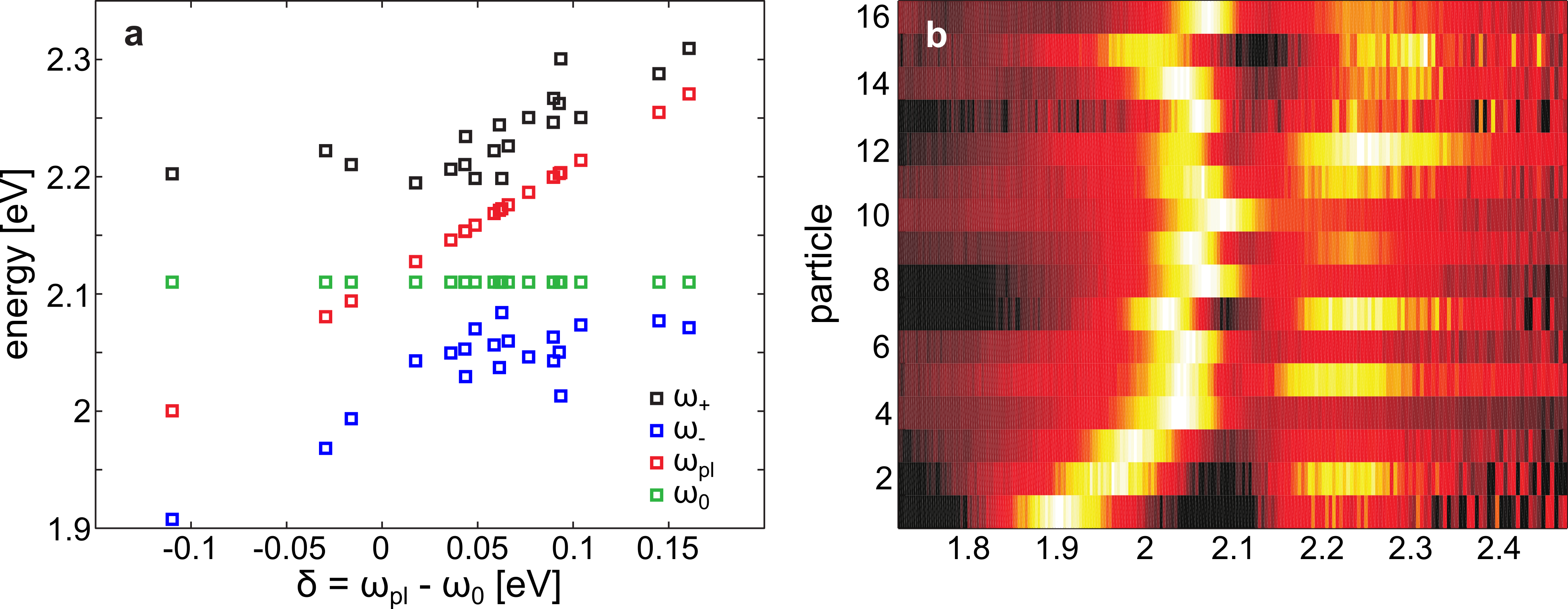}
\caption{(a) Hybridized mode frequencies ($\omega_+$,$\omega_-$) plotted against the detuning $\delta$. The plot resembles anti-crossing behavior with average vacuum Rabi splitting of $\Omega_{R}\approx170$~meV. (b) Scattering spectra for all studied nanoparticles, ordered according to detuning.}
\label{fig::S5}
\end{figure}

\newpage

\section{\textbf{S6}: Numerical calculations of electromagnetic near fields, mode volumes and quality factors for uncovered silver nanoprisms}

\begin{figure}[h]
\centering
\includegraphics[width=12.5cm]{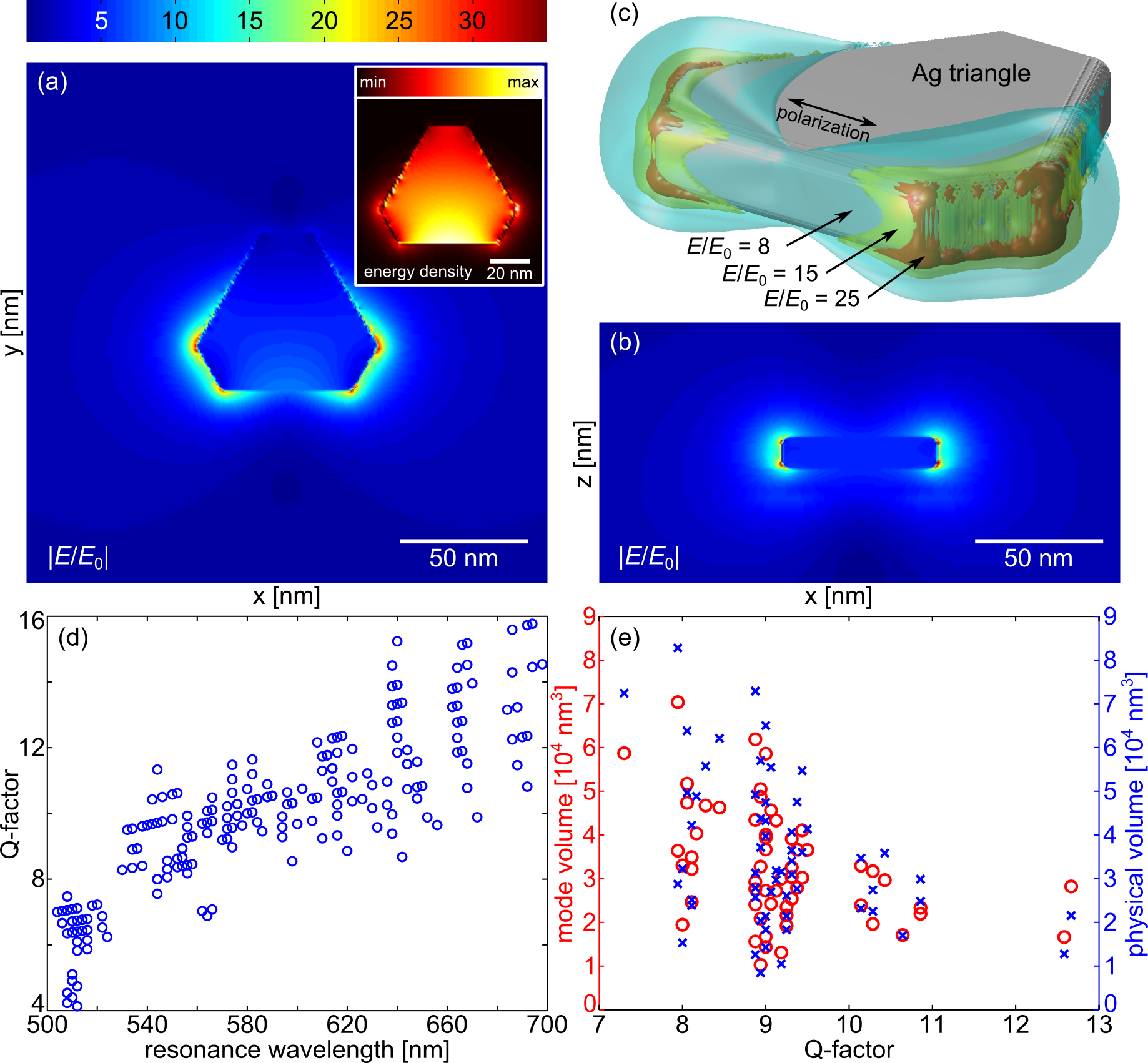}
\caption{Optical properties of bare silver truncated triangles on formvar illuminated by x-polarized light (thickness $t$, longer edge $b$, shorter edge $a$). (a) Electric near-field in a plane through the center of the triangle ($t=12$~nm, $b=50$~nm, $a=20$~nm) and (b) an xz-cross-section through the widest part of the triangle. The color scale is the same for (a,b) and shows the field amplitude. The inset in (a) shows a cross section of the energy density in a silver nanotriangle. Note, that most of the energy is contained within the nanoparticle itself. (c) A 3D visualization of the same triangle with marked regions that have an electric field enhancement of at least 8 (cyan), 15 (yellow), and 25 (orange). (d) Q-factor vs. resonance wavelength of silver triangles of sizes $t=10\div20$~nm, $b=20\div100$~nm, $a=5\div25$~nm. (e) Mode volume (left y-axis, red circles) and physical volume (right y-axis, blue crosses) vs. Q-factor for triangles with resonance wavelengths $\pm20$ nm from the J-aggregate line of 
588 nm. The simulated triangles are chosen to map the span of dimensions observed in experiments.}
\label{fig::S6}
\end{figure}

\end{widetext}

\subparagraph{Mode volume calculation} 

The mode volume is calculated using
\begin{equation}
V=\frac{\int\epsilon(r)|E(r)|^2 \ud V}{\mathrm{max}(\epsilon(r)|E(r)|^2)},
\end{equation}
however, with the modification to make the equation applicable to plasmons, where the term $\epsilon(r)\to\mathrm{Re}[\epsilon]+2\omega\mathrm{Im}[\epsilon]/\gamma$, where $\epsilon$ is the permittivity of the metal, $\gamma$ is the Drude damping term, and Re[] and Im[] are, respectively, the real and imaginary parts \cite{PhysLettA_299_309_ruppin, OL_35_4208_koenderink}.

Calculating the mode volume from FDTD simulations may encounter problems that result from staircase meshing which may give very large field enhancements over volume spanning one or at most a few pixels (nodes). This makes the denominator in the mode volume equation very large and make $V$ abnormally small. To mitigate this effect, we analyze the energy distributions in a number of nanotriangles and identify locations where it assumes maximum values due to physical, not numerical, effects. For triangles illuminated by linearly polarized light as in simulations the maximum energy density occurs along the long edge parallel to the polarization, see inset in Fig. S6a. 

\newpage

\section{\textbf{S7}: Numerical calculations of scattering and absorption cross-sections of nanoprisms interacting with J-aggregate sheets}

\begin{figure}[h]
\centering
\includegraphics[width=8.5cm]{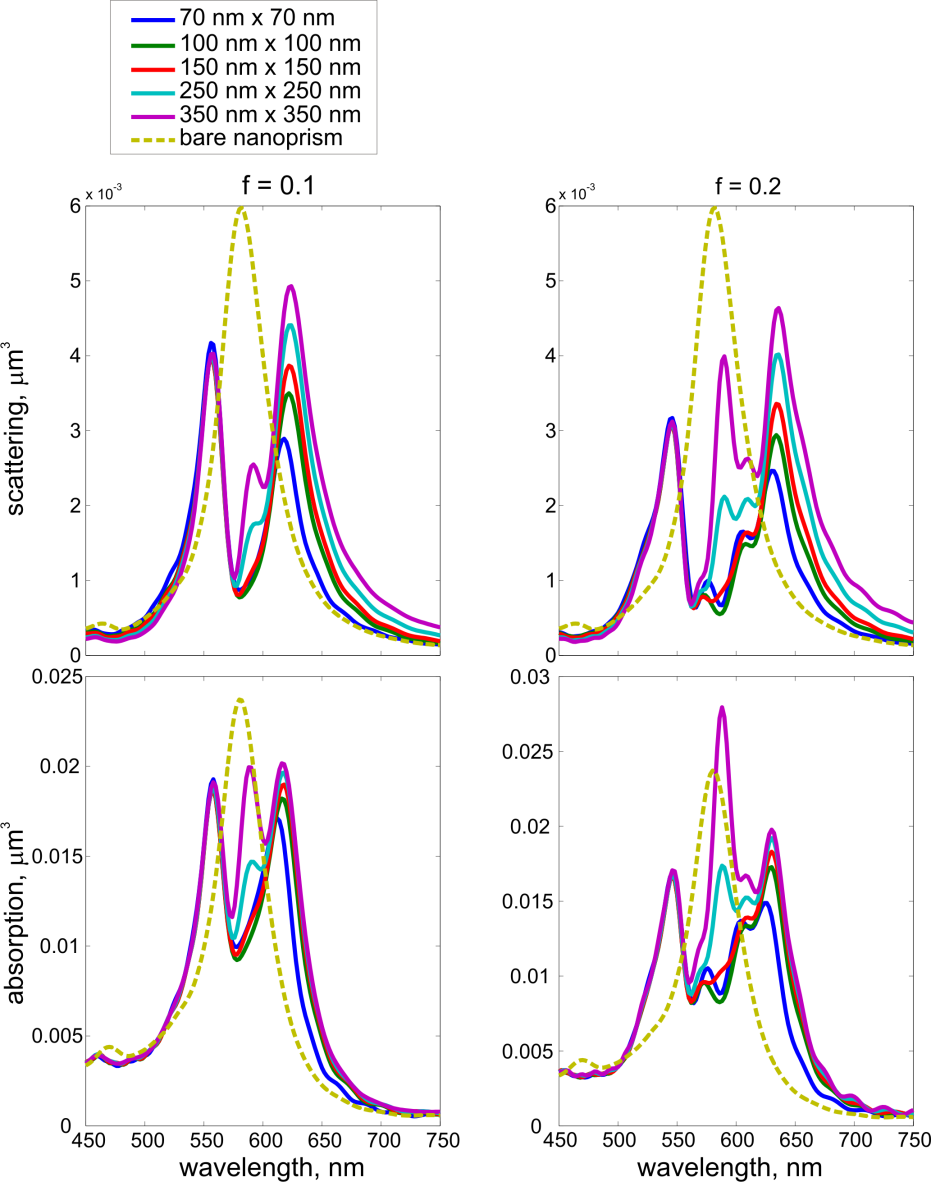}
\caption{Scattering and absorption cross-sections. Scattering and absorption cross-sections of a triangle from Fig. S6 ($t = 12$~nm, $b = 50$~nm, $a = 20$~nm) placed on a sheet of J-aggregate on formvar. Top row shows the scattering cross-section, bottom row the absorption cross-section, while columns feature increasing oscillator strengths f.}
\label{fig::S7}
\end{figure}

Note, that for the investigated triangles their absorption cross section is about 5 times larger than scattering. Thus, scattering spectra feature much lower dips due to coupling between the plasmon and the J-aggregate than absorption spectra. Solid lines indicate coupled spectra for Ag triangles on a single rectangular sheet of J-aggregate ranging from 70 by 70~nm to 350 by 350~nm. Note, that very large sheets of J-aggregate with large $f$ should mark their presence by narrow peaks at the absorption line -- these originate from molecules uncoupled or weakly coupled to the silver triangle. The upper limit of the oscillator strength is estimated based on the absorption cross-section of one exciton (15 molecules). For simplicity we assume a spherical structure with a Lorentzian line (50 meV linewidth) whose peak absorption we vary by changing the oscillator strength, with the volume given by the geometrical volume of 15~TDBC molecules (the volume of one TDBC molecule is approximately 0.5~nm$^3$.). To match the 
literature reported value of $2\times10^{-14}$~cm$^{2}$ we need to set $f = 0.5$ with a sphere 7.5~nm$^3$ in volume. However, this yields large negative permittivity values on the order of -10 (real part) and gives spectra not observed experimentally. In order to reliably match the measurements we need to reduce f (i.e. dilute the J-aggregate). Very good agreement is achieved for oscillator strengths on the order of $0.1\div0.15$. Assuming $f = 0.1$ the effective volume of one molecule has to be 2.5~nm$^3$, meaning that the estimated 1000 TDBC molecules ($70\div85$ excitons) coupled to the silver triangle occupy about 2500~nm$^3$. If we further take that the sheets are approximately 3~nm thick, this means that the active molecules occupy a surface area on the order of 830~nm$^2$, a value comparable or smaller to the geometrical cross section of a typical triangle. Assuming unpolarized light, the hot-spots around the triangle are concentrated at the corners (or narrower sides for a truncated triangle as in 
Fig. S6, though there we only show one polarization) and the total length of those corners and edges is about 220~nm. The fields extend laterally (with respect to the circumference) for at least a few nanometers, meaning that the area exposed to strongly enhanced near-fields is enough to fit a considerable amount of the estimated 1000~TDBC molecules. 

\newpage

\begin{widetext}
\section{\textbf{S8}: Fluorescence data for five exemplary nanoparticles with diverse coupling}

\begin{figure}[h]
\centering
\includegraphics[width=12cm]{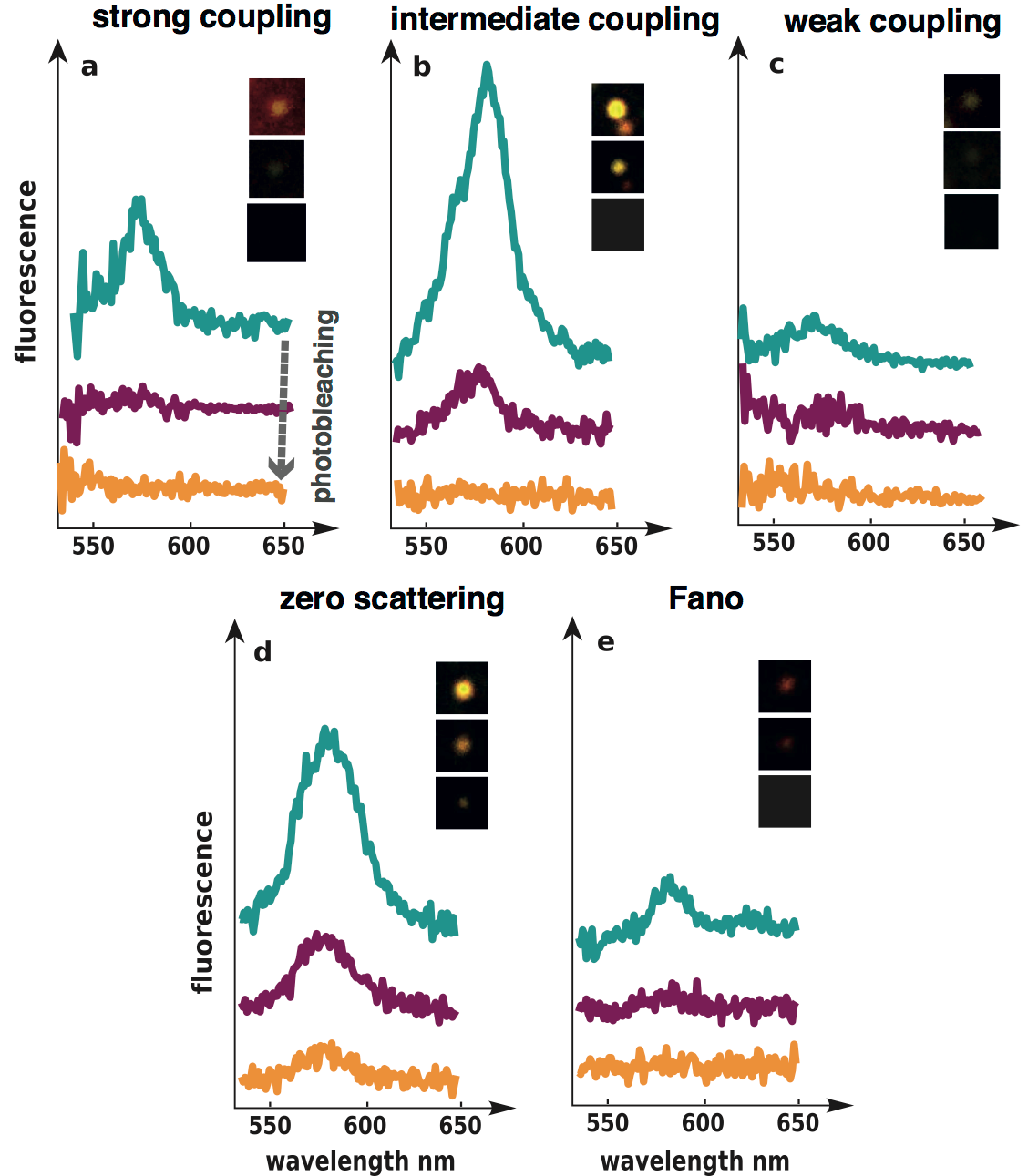}
\caption{Fluorescence spectra as a function of photobleaching time for five notable cases shown in Figure 3 of the main text. Inset: color fluorescence images corresponding to the three spectra (image dimensions correspond to 2.5~$\mu$m). Note that no fluorescence signal from bare silver nanoprisms could be detected, indicating that fluorescence originates exclusively from the adsorbed dye molecules.}
\label{fig::S9}
\end{figure}
\end{widetext}

Fluorescence of free TDBC J-aggregates in water solution exhibits a narrow (fwhm $\sim35$~meV) spectrum peaking at 588 nm, i.e. with nearly zero Stokes shift \cite{SciRep_3_3074_gulis}. In contrast, fluorescence of J-aggregates interacting with silver nanoprisms is much broader, up to $\sim175$~meV as in Fig. S8a, indicating strong interaction between plasmons and excitons. Similarly to photodegradation of scattering spectra observed in Fig. 3 of the main text, fluorescence of coupled J-aggregates also decays as a function of time, confirming it originates from the J-aggregates.

\begin{widetext}
\newpage
\section{\textbf{S9}: Control measurements on bare Ag nanoprisms}

\begin{figure}[h]
\centering
\includegraphics[width=15cm]{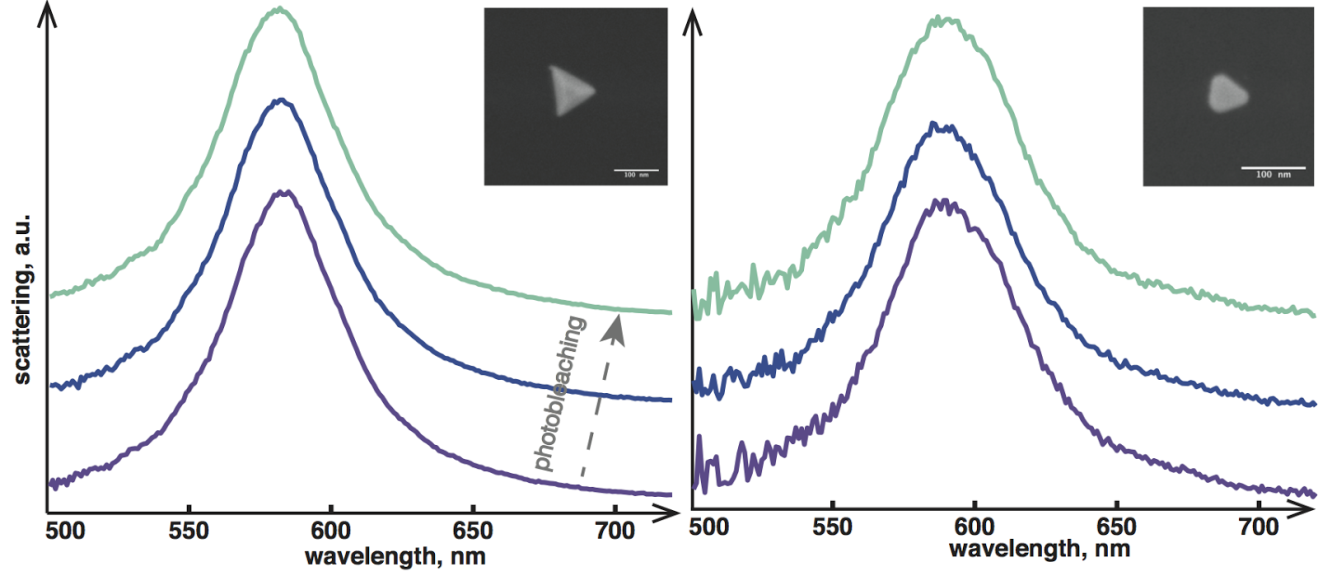}
\caption{Dark-field scattering spectra for bare triangular shaped Ag nanoparticles after no (purple), 2 min (blue) and 10 min (cyan) of laser illumination. The experiments were performed under the same condition as J-aggregate photobleaching. No observable effect of laser illumination on particle spectra was detected. No inelastic scattering e.g. fluorescence could be detected from bare nanoprisms under the experimental conditions used.}
\label{fig::S10}
\end{figure}

\end{widetext}


%

\end{document}